\documentclass[10pt,journal]{IEEEtran}
\usepackage{amsmath}
\usepackage{epsfig}
\usepackage{balance}
\usepackage[noadjust]{cite}
\newtheorem{lemma}{Lemma}[section]

\newcommand{\qed}{\nobreak \ifvmode \relax \else
      \ifdim\lastskip<1.5em \hskip-\lastskip
      \hskip1.5em plus0em minus0.25em \fi \nobreak
      \vrule height0.75em width0.5em depth0.25em\fi}

\begin{document}
\title{A New Non-Iterative Decoding Algorithm for the Erasure Channel
: Comparisons with Enhanced Iterative Methods}
\author{
\authorblockN{J. Cai, C. Tjhai, M. Tomlinson, M. Ambroze and M. Ahmed}\\
\authorblockA{Fixed and Mobile Communications Research\\
University of Plymouth\\
PL4 8AA, United Kingdom\\
email: \{jcai,ctjhai,mtomlinson,mambroze,mahmed\}@plymouth.ac.uk}}

\maketitle
\begin{abstract}
This paper investigates decoding of binary linear block codes over
the binary erasure channel (BEC). Of the current
iterative decoding algorithms on this channel,
 we review the Recovery Algorithm and
the Guess Algorithm. We then present a Multi-Guess Algorithm extended
from the  Guess Algorithm and a new algorithm -- the In-place Algorithm.
The Multi-Guess Algorithm can push the limit to break the stopping sets.
However, the performance of the Guess and the Multi-Guess Algorithm depend on the parity-check 
matrix of the code. 
Simulations show that we can decrease
the frame error rate by several orders of magnitude using the Guess and
the Multi-Guess Algorithms when the parity-check matrix of the code is 
sparse. The In-place Algorithm can obtain better performance 
even if the parity check matrix is dense. 
We consider the application of  these algorithms in the implementation of 
 multicast and broadcast techniques on the Internet.
 Using these algorithms, a user does not have to wait 
until the entire transmission has been received.
\end{abstract}

\section{Introduction}

The Binary Erasure Channel (BEC) was introduced by Elias \cite{Elias} in 1955. It counts lost 
information bits as being ``erased'' with probabilities equal to $0.5$. Currently, the BEC is widely
used to model the Internet transmission systems, in particular multicasting and broadcasting.

As a milestone, Luby {\em et. al.} \cite{Luby.1} proposed the first realization of a class of 
erasure codes -- LT codes, which are rateless and are generated on the fly as needed. However, LT-codes cannot 
be encoded with constant cost if the number of collected output symbols is close to the number of input 
symbols. In \cite{Shokrollahi}, Shokrollahi introduced the idea of Raptor codes which adds an outer
code to LT codes. Raptor codes have been established in order to solve the error floors exhibited by the
LT codes.

On the other hand, low-density parity-check (LDPC) codes have been studied \cite{Luby.2} to \cite{C.Di} for application to the BEC.
 The iterative decoding algorithm, which is the same as Gallager's soft-decoding algorithm \cite{Gallager},
was implemented \cite{Luby.2}. Capacity-achieving degree distributions for the binary erasure channel
have been introduced in \cite{Luby.2}, \cite{Shokrollahi.2} and \cite{Oswald}. Finite-length analysis of LDPC
codes over the BEC was accomplished in \cite{C.Di}. In that paper, the authors
have proposed to use finite-length
analysis to find good finite-length codes for the BEC. 

In this paper, we show the derivation of a new decoding algorithm to improve 
the performance of binary linear block codes on the BEC. The algorithm can
be applied to any linear block code and is not limited to LDPC codes.
Starting with superposition of the erased bits on the parity-check matrix, we 
review the performance of the iterative decoding
algorithms, described in the literature, for the BEC, principally the Recovery
 Algorithm and 
 the Guess Algorithm~\cite{H.P}. In Section \ref{sec:03}, we propose an improvement
to the Guess Algorithm based on multiple guesses: the Multi-Guess Algorithm 
and give a method to calculate the minimum number of guesses required in the decoding procedure. 
In this section, 
we also describe a new, non iterative decoding algorithm based
 on a Gaussian-Reduction method \cite{Odlyzko} by
 processing the parity-check matrix. In Section \ref{sec:04}, 
we compare the performance of these algorithms for different codes 
using computer simulation.
 In Section \ref{sec:05}, we discuss the application of these decoding algorithms for the Internet.
Section \ref{sec:06} concludes the paper. 

\section{Preliminaries}\label{sec:02}
\subsection{Matrix Representations of the Erased Bits}
Let $H$ denote the parity-check matrix.
 Considering an $L\times N$ binary linear block code, we assume that the encoded sequence is $\mathbf{x} = \{x_1, x_2, \ldots, x_N\}$.
After being transmitted over the erasure channel with erasure probability $\epsilon$, the encoded 
sequence can be divided into the transmitted sub-sequence and the erased sub-sequence, denoted as
$\mathbf{y} = \{y_1, y_2,\ldots, y_{l_1}\}$
 and $\mathbf{y_{\epsilon}} = \{y_{{\epsilon}1},
 y_{{\epsilon}2}, \ldots, y_{{\epsilon}l_2}\}$
respectively, where $l_1 + l_2 = N$.

\begin{figure}[hbt]
\centering
\includegraphics[width=1.5in]{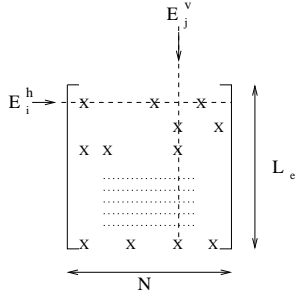}
\caption{\label{fig:1}A matrix representation of the erased bits}
\end{figure}

Corresponding to the parity check matrix of the code, we can generate an erasure matrix 
$M_{\epsilon}$ ($L_\epsilon \times N$) which contains the
positions of the erased bits in $H$.
Then we denote the set of erased
bits $i$ that participate in each parity 
check row by $E_i^h = \{j:M_{\epsilon(ij)}=1\}$ with $h$ standing for 
``horizontal'' and the number of erased 
bits in $E_i^h$ is denoted by $|E_i^h|$.
Similarly we define the set of checks in which bit $j$ participates, $E_j^v=
\{i:M_{\epsilon(ij)}=1\}$ with $v$ standing for ``vertical'',
 and the number of erased bits in $E_j^v$
 is denoted by $|E_j^v|$.
 Let $\mathbf{E^h} = \{E_i^h\ |\ i \in \{1, 2, \ldots,
L_\epsilon\}\}$ and $\mathbf{E^v} = \{E_j^v\ |\ j \in \{1, 2, \ldots,
N\}\}$. The matrix representation is shown in Fig. \ref{fig:1}, where
an ``x'' represents an erasure.

\subsection{Current Iterative Decoding Algorithms for the BEC}

In \cite{Luby.2}, the message-passing algorithm was used for 
reliable communication over the BEC at 
transmission rates arbitrarily close to channel capacity. The decoding 
algorithm succeeds if and only if the set of erasures do not cause 
stopping sets~\cite{C.Di}. For completeness, this algorithm is briefly
outlined below:\\
{\bf Recovery Algorithm}
	\begin{itemize}
		\item {\em step 1} Generate the $M_{\epsilon}$ and 
obtain the $\mathbf{E^h}$.
		\item {\em step 2} For $i \in \{1, 2, \ldots, L_{\epsilon}\}$,
 if $|E_i^h| = 1$, we replace the value in the bit position $i$ with 
the XOR of the unerased bits in that check equation.
 Then we remove the erasure from the erasure matrix.
		\item {\em step 3} Continue from step 2 until all the erased bits
 are solved or the decoding cannot continue further.
	\end{itemize}
%

The decoder will fail if stopping sets exist. 

We can break the stopping sets
by performing several ``guesses'' of the unsolved erased bits. This algorithm
is called the Guess Algorithm~\cite{H.P}.
\\
{\bf Guess Algorithm}
	\begin{itemize}
		\item {\em Step 1} Run the decoder with Recovery Algorithm
 until it fails due to stopping set(s).
		\item {\em Step 2} In order to break the stopping set,
 when $|E^h_i| = 2$, we guess one of the erased symbols and update the erasure matrix $M_{\epsilon}$ 
and $\mathbf{E^h}$.
		\item {\em Step 3} Continue from step 1 until
 all the erased symbols are solved or the decoding cannot continue further.
	If the decoder cannot continue, declare a decoder failure and
exit.
		\item{\em Step 4} Creat a list of $2^g$ solutions, where
$g$ is the number of guesses made. From the list $c_{\text{out}_k}$
, $k \in \{1, 2, \ldots, 2^{g}\}$, pick the one that satisfies 
$Hc^{T}_{\text{out}_k} = \mathbf{0}$.
	\end{itemize}
%
%

Obviously, compared to the Recovery Algorithm, the complexity of
this algorithm increases with $g$.
 Usually, we limit the number of guesses
to a small number $g_{\text{s}}$. If after $g_{\text{s}}$ guesses, the decoding
still cannot be finished, a decoding failure is declared.
For sparse codes with low-density $H$, e. g. LDPC codes, the Guess Algorithm 
can improve the performance with $g < 3$ guesses as shown in Section \ref{sec:04}.

The decoding algorithm is more efficient when the bits to be 
guessed are carefully chosen. These are termed ``crucial''bits.
The crucial bits are chosen on the basis of the highest value of
 $|E^v_j|$ with the value of $|E^h_j| = 2$. 
\section{Improved Decoding Algorithms for Non-sparse Linear Block Codes for 
the BEC}\label{sec:03}
For non-sparse linear codes, it is common to encounter more than 2
unsolved symbols in each row of $M_{\epsilon}$ after
 running the Guess Algorithm, 
due to the high-density of their parity check matrix. In these cases,
 we cannot break the stopping set by guessing one erased symbol in a row only.
 More than 1 erased symbols at one time need to
be guessed. We can calculate the minimum number of 
guesses before the decoding.

\begin{lemma}
Consider the chosen erased symbols in each row as an erased group.
Let $\omega_{\delta}$ denote the set of rows with $\delta$ erasures, that
is, $\omega_{\delta} = \{i\ |\ |E_i^h|=\delta\}$. And $x_{\delta}$
is the set of rows which satisfies:\\
\begin{equation}
	x_{\delta} = \{i \in \omega_{\delta}\ |\ \exists k, p \in E_i^h, 
\text{such as}\ k\neq p, |E_k^v|=|E_p^v|=1\}.
\end{equation}
Then 
\begin{equation}
\min{g}=|x_{\delta}|+1
\end{equation}
 where 1 accounts for the need for at least one ``crucial'' row.
\end{lemma}
\begin{proof}
	When the guessing process stops, there are more than 
2 erased symbols in each erased row.
The rows that have more than two bits $(k, p)$ which do 
not participate in any other row (i. e. $|E_k^v| = |E_p^v| = 1$)
 cannot be solved by other rows, and so at least one of these
bits has to be guessed.
So the minimum number of guesses equals to the number of all the independent
guesses plus one more ``crucial'' guess to solve the other rows.
\end{proof} 
For the Multi-Guess Algorithm, a whole row is guessed. A crucial
row $c$ is defined as follows:

\begin{enumerate}
	\item $c \in \omega_{\delta}$
	\item $\sum_{j \in E_c^h} |E_j^v|\  \text{is maximized over}\ c\ \text{in} 
\ \omega_{\delta}$
\end{enumerate}

The Multi-Guess Algorithm is given below:\\
{\bf Multi-Guess Algorithm}
	\begin{itemize}
		\item {\em step 1} Run the decoder with Guess
Algorithm until $|E_i^h| > 2$ for $i = 1, \ldots, L_{\epsilon}$.
		\item {\em step 2} Evaluate the value of $\min(g)$. If
$\min(g) > g_{\text{s}}$, the decoding declares a failure and
exits. 
		\item {\em step 3} Group the rows with $|E^h_i| = \delta$
as $\omega_{\delta}$, where $i \in \{1, 2, \ldots, L_{\epsilon}\}$. 
		\item {\em step 4} Find the ``crucial'' row and guess
all erased bits in that row. (There will be at most $2^{\delta-1}$ guesses.)
		\item {\em step 5} Guess one bit $p$ with
$|E_p^v|=1$ in each of the independent rows, i.e.
the rows in $x_g$.
		\item {\em step 6} Update $M_{\epsilon}$, $\mathbf{E^h}$ and 
$\mathbf{E^v}$. Continue the decoding from step 3 
to step 5 until all the erased bits are solved or the decoding cannot 
continue further.
	\end{itemize}

The disadvantages of Guess and Multi-Guess Algorithms include
 the decoding complexity and the correctness of the results. 
The decoding complexity grows exponentially with the number of guesses.
It is possible that the group guess declares a wrong value 
as the result of the decoder. Although this kind of situation happens only
 when the value of $\epsilon$ is very small, it is still undesirable.

Let $\mathbf{x'}$ denote the received vector, where $\mathbf{x'} = 
\mathbf{y} \bigcup \mathbf{y_{\epsilon}}$. We now devise a reduced
complexity algorithm to decode the erased bits by solving the 
 equation~\ref{eq:1} using the Gaussian Reduction method \cite{Odlyzko}.
\begin{equation} \label{eq:1}
	H\mathbf{x'^{T}} = \mathbf{0}.
\end{equation}
According to \cite{C.Di}, the optimal decoding is equivalent to solving 
the linear system, shown in the equation \ref{eq:1}. If the equation \ref{eq:1}
has a unique solution, the optimal algorithm is possible.
Guassian Reduction algorithm is considered as the optimal algorithm
 over the BEC. We propose a reduced complexity Guassian Reduction algorithm
-- In-place Algorithm \cite{patent} by elimilating the column-permutations required.
This algorithm is stated as follows:
\\
{\bf In-place Algorithm}
\begin{itemize}
	\item{\em step 1}  The codeword is received
and $y_{\epsilon}$ are substituted in positions of erased bits in 
$H$. Starting with one of the erased symbols, $y_{\epsilon_{s}}$, the first
 equation containing this symbol is flagged that it will be used for
the solution of $y_{\epsilon_{s}}$. This equation is subtracted from 
all other equations containing $y_{\epsilon_{s}}$ and not yet flagged to produce
a new set of equations. The procedure repeats until either non flagged
equations remain containing $y_{\epsilon_{s}}$ 
(in which case a decoder failure is declared) or no erased symbols remain that 
are not in flagged equations.
    \item{\em step 2} Let $y_{\epsilon_{\text{last}}}$ be the erased
symbols at the last flagged equations. In the latter case, starting with 
$y_{\epsilon_{\text{last}}}$ this equation is solved to find $y_{\text{last}}$
 and this equation is unflagged. This coefficient is
substituted back into the remaining flagged equations containing
 $y_{\text{last}}$.
The procedure now repeats with the second from last flagged eqaution now 
being solved for $y_{\epsilon_{\text{last}-1}}$. This equation is unflagged and
followed by back substitution of $y_{\text{last}-1}$ for 
$y_{\epsilon_{\text{last}-1}}$
in the remaining flagged equations.
\end{itemize}

\begin{figure}[hbt]
\centering
\includegraphics[width=2.5in,angle=0]{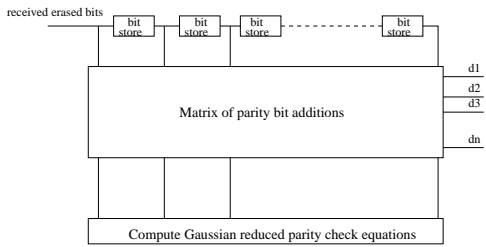}
\caption{\label{fig:03}Erasure Correction Using In-place Algorithm}
\end{figure}

A block schematic of the decoder is shown in Fig.\ref{fig:03}.
 The received bits are 
stored in the shift register with the erased bits being replaced by the 
unknown $y_{\epsilon}$. The Gaussian reduced equations are computed
and used to define the connection of bit adders from the respective shift
register stage to compute the outputs $d_1$ to $d_n$. The 
non erased symbols contained in the shift register are switched directly 
through to the respective output so that the decoded codeword with no
erased bits is present at the outputs $d_1$ through to $d_n$.
\\

\section{Results}\label{sec:04}
\begin{figure}[hbt]
\centering
\includegraphics[width=2.5in,angle=0]{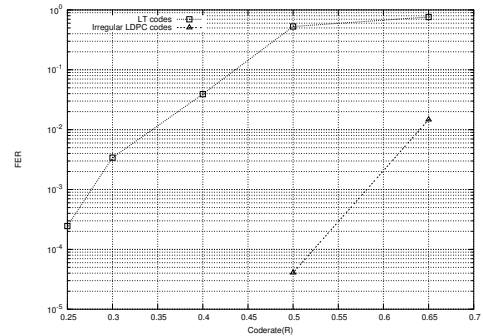}
\caption{\label{fig:04}Performance of the LT codes and irregular LDPC codes with erasure
probability = 0.2}
\end{figure}

We evaluated the performance of the Recovery Algorithm with the LT codes with
Soliton distribution as described in \cite{Luby.1} and irregular LDPC codes. As shown in Fig. \ref{fig:04},
the performance of irregular LDPC codes is significantly better than that of the LT codes
for the same block length. As a consequance, we use LDPC codes to benchmark
the remaining algorithms.

A particularly strong binary code and which has a sparse $H$ is
the cyclic LDPC code (255,175), which has a length of 255 bits after encoding
of 175 information bits. Since the parity-check polynomial of the (255,175)
\footnote{The (255,175) Cyclic LDPC code has a minimum Hamming distance of 17.}
code is orthogonal on every bit position, the minimum Hamming distance is 
$1+w$, where $w$ denotes the number of ones per row in $H$~\cite{peterson}. 

\begin{figure}[hbt]
\centering
\includegraphics[width=2.5in,angle=0]{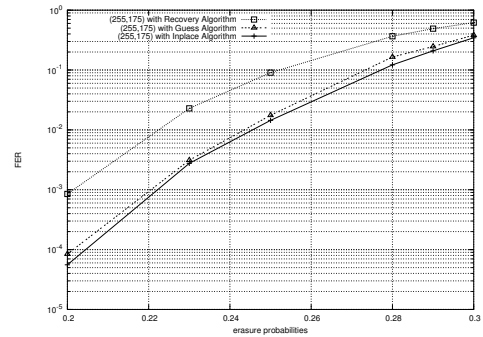}
\caption{\label{fig:06}Performance of the Cyclic LDPC (255,175) with the Guess 
, the Multi-Guess and the In-place Algorithms}
\end{figure}

The applicability of the decoding methods above depends on the error 
correcting code being used and specifically on the parity check matrix being used. The
 performance of this code for the Recovery,
the Guess and the In-place Algorithms is shown in Fig. \ref{fig:06} in terms of
the probability of decoder error (FER) as a function of the erasure probability
for every transmitted bit. 

\begin{figure}[hbt]
\centering
\includegraphics[width=2.5in,angle=0]{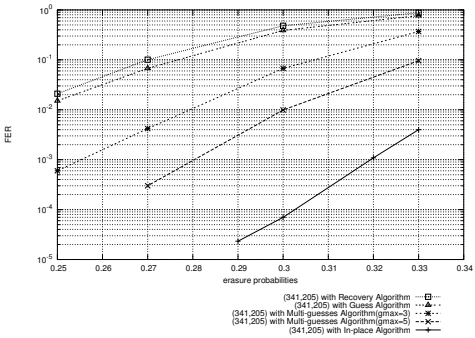}
\caption{\label{fig:07}Performance of the Cyclic LDPC (341,205) 
with the Recovery, the Guess, the Multi-Guess and the In-place Algorithms}
\end{figure}

Due to its sparse parity check matrix, Guess algorithm with less than 3 
guesses can achieve more than 1 order of magnitude improvement compared to that
of Recovery Algorithm. In addition, from Fig. \ref{fig:06}, we also can see 
that the curve of Guess Algorithm is very close to the curve of
 In-place Algorithm, which means Guess Algorithm is a ``near optimal decoding''
algorithm when it has a sparse parity check matrix.

Fig. \ref{fig:07} shows the performance of the (341,205) LDPC code
\footnote{The (341,105) LDPC code has a minimum Hamming distance of 16.} with 
the Recovery, the Guess, the Multi-Guess and the In-place Algorithms. Comparing these results of the Recovery and Guess Algorithms,
 the Multi-Guess Algorithm can obtain the results by several orders 
of magnitude better. For example, when the erasure probability equals to 0.3, 
the Multi-Guess Algorithm with $g_{\text{max}}=3$ is one order of magnitude
 better than the Recovery and Guess Algorithms, when $g_{\text{max}}=5$, 
the Multi-Guess Algorithm is 2 order2 of magnitude better than the Recovery
and the Guess Algorithms. As an optimal decoding algorithm, the In-place Algorithm can achieve 4 orders of magnitude 
better than the Recovery and the Guess Algorithm.

\begin{figure}[hbt]
\centering
\includegraphics[width=2.5in,angle=0]{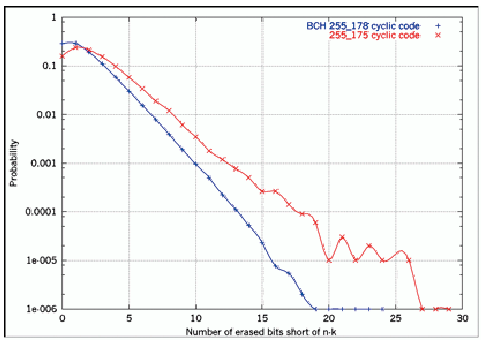}
\caption{\label{fig:6} Comparison of Probability Distribution of 
Number of Erased Bits not Corrected from Maximum Correctible (N-L) for 
(255,175) code and BCH (255,178) code}
\end{figure}

The ultimate performance of the In-place Algorithm as a function of error correcting
code is shown in Fig. \ref{fig:6} for the example (255,175) code which can
correct a maximum of 80 erased bits.  Fig. \ref{fig:6} shows
the probability density function of the number of erased bits
short of the maximum correctable which is $N-L$.
 The results were obtained by computer simulations.
The probability of being able to correct only 68 bits, a shortfall of 12 bits,
is $1.1\times10^{-3}$. Simulations indicate that on average 77.6 erased bits
may be corrected for this code. In comparison the BCH (255,178) code having 
similar rate is also shown in Fig. \ref{fig:6}. The BCH code has a similar
rate but a higher minimum Hamming distance of 22 (compared to 17). It can 
be seen that it has better performance than the (255,175) code but it
has a less sparse parity check matrix and consequently it is less suitable
for Recovery Algorithm and Guess Algorithm. Moreover the average shortfall 
in erasures not corrected is virtually identical for the two codes.

\begin{figure}[hbt]
\centering
\includegraphics[width=2.5in,angle=0]{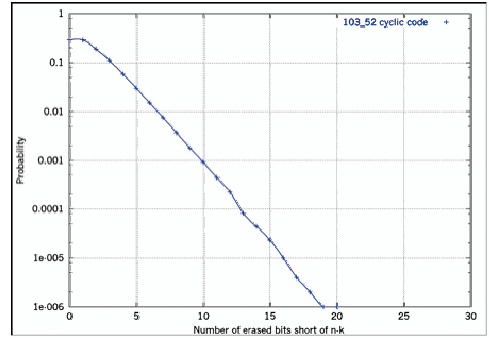}
\caption{\label{fig:7} Probability Distribution of Number of Erased
Bits not Corrected from Maximum Correctible (N-L) for 
(103,52) code quadratic redisue code}
\end{figure}

The simulation results of using In-place Algorithm for the (103,52) 
quadratic residue binary code \cite{F.J} are shown in Fig. \ref{fig:7}.
The minimum Hamming distance for this code is 19 and the results
are similar to that of the (255,178) BCH code above. It is found
from the simulations that on average 49.1 erasure bits are corrected
(out of a maximum of 51) and the average shortfall from the 
maximum is 1.59 bits.

\begin{figure}[hbt]
\centering
\includegraphics[width=2.5in,angle=0]{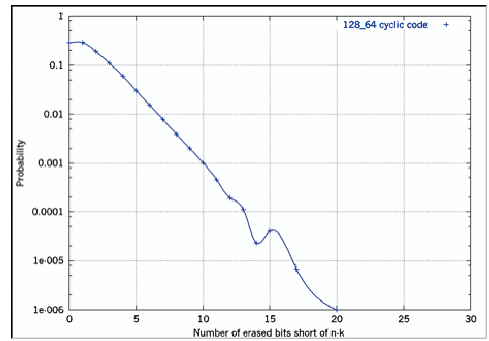}
\caption{\label{fig:8} Probability Distribution of Number of Erased 
Bits not Corrected from Maximum Correctible (N-L) for 
(128,64) extended BCH code}
\end{figure}

Similarly the results for the extended BCH (128,64) code is shown
in Fig. \ref{fig:8}. This code has a minimum Hamming distance of 22
and has a similar probability density function to the other BCH codes
above. On average 62.39 erasure bits are corrected (out of a maximum
of 64) and the average shortfall is 1.61 bits from the maximum.

\section{Application}\label{sec:05}

In multicast and broadcast information is transmitted in data
packets with typical lengths from 30 bits to 1000 bits. These
packets could define a symbol from a Galois field \cite{peterson},
viz $GF(2^m)$ but with $m$ equal to 30 or more up to and beyond
1000 bits this is impracticable and it is more convenient to use a 
matrix approach with the packets forming the rows of the matrix 
and columns of bits encoded using an error correcting code. Usually,
but not essentially the same code would be used to encode each 
column of symbols. The matrix of symbols may be defined as:

\begin{table}[hbt]
	\centering
		\begin{tabular}{l l l}
			$b_{00}b_{01}b_{02}b_{03} \ldots b_{0s}$&$=$&packet 1\\
			$b_{10}b_{11}b_{12}b_{13} \ldots b_{1s}$&$=$&packet 2\\
			$b_{20}b_{21}b_{22}b_{23} \ldots b_{2s}$&$=$&packet 3\\
			$\cdots \cdots \cdots \cdots \cdots$&$\cdots$&$\cdots \cdots$\\	
			$\cdots \cdots \cdots \cdots \cdots$&$\cdots$&$\cdots \cdots$\\	
			$\cdots \cdots \cdots \cdots \cdots$&$\cdots$&$\cdots \cdots$\\	
			$b_{n-10}b_{n-11}b_{n-12}b_{n-13} \ldots b_{n-1s}$&$=$&packet n
		\end{tabular}
\end{table}

There are a total of $(s+1)\cdot k$ information symbols which
encoded using the parity check equations of a selected code into 
a total number of transmitted symbols equal to $(s+1)\cdot n$. The
symbols are transmitted in a series of packets with each packet
corresponding to a row as indicated above. For example the row:
$b_{20}b_{21}b_{22}b_{23}\ldots \ldots b_{2s}$ is transmitted as
a single packet.

Self contained codewords are encoded from each column of symbols. 
For example $b_{00}b_{10}b_{20}\ldots \ldots b_{k-10}$ form the
information symbols of one codeword and the remaining symbols, 
$b_{k+0}b_{k+10}b_{k+20}\ldots \ldots b_{n-10}$ are the
parity symbols of that codeword. As a result of network congestion,
drop outs, loss of radio links or other multifarious reasons not 
all of the transmitted packets are received. The effect is that 
some rows above are erased. The decoding procedure is that codewords
are assemble from the received packets with missing symbols corresponding
to missing packets marked as $z_{ij}$. For example, if the second 
packet only is missing above:
\begin{itemize}
	\item The first received codeword corresponds to the first
column above 
 and is $b_{00}z_{10}b_{20}\ldots \ldots b_{n-10}$
	\item The second codeword corresponding to the first column 
above 
and is $b_{01}z_{11}b_{21}\ldots \ldots b_{n-11}$ and so on.
\end{itemize}

All the algorithms stated in Section 2 may be used to solve for 
the erased symbols $z_{10}$ in the first received codeword,
and for the erased symbol $z_{11}$ in the second received 
codeword and so on up to the $s'$th codeword (column) solving
for $z_{1s-1}$.
\balance
 
As an example, the binary, extended $(128,64)$ BCH code could
be used to encode the information data. The packet length is chosen
to be 100 bits, and the total transmission could consist of 128 transmitted
packets (12,800 bits total) containing 6,400 bits of information. On
average as soon as any 66 packets from the original 128 packets have
been received, the remaining 62 packets are treated as if they
are erased. 100 codewords are assembled, decoded with the erasures
solved and the 6,400 bits of information retrieved. One advantage is 
that a user does not have to wait until the entire transmission has
been received.

\section{Conclusions}\label{sec:06}
In this paper, we presented different decoding algorithms of LDPC
codes over the BEC: Recovery, Guess, Multi-Guess and In-place Algorithms.
The Multi-Guess Algorithm is an extension to Guess Algorithm, which
can push the limit to break the stopping sets. 
We show that Guess and Multi-Guess Algorithms are parity-check matrix 
dependent. For the codes with sparse parity-check matrix, Guess and
Multi-Guess Algorithms can be considered as ``Near-optimal Decoding 
Methods''. On the other hand, In-place Algorithm is not. It's an
optimal method for the BEC and able to correct $N - L - \rho$ erasures,
where $\rho$ is a small positive integer.

We also considered these algorithms in the implementation of 
 multicast and broadcast. Using these algorithms, a user does not have to wait 
until the entire transmission has been received.

\end{document}